\documentclass[11pt,reqno]{amsart} 
\usepackage{amsmath,amsgen,amstext,amsbsy,amsopn,amsthm,amssymb} 
 
\newtheorem{thm}{THEOREM} 
 
\newtheorem{lem}{LEMMA} 
 
\theoremstyle{definition} 
\newtheorem{rem}{REMARK}

\newcommand{\infspec}{{\rm inf\ spec\ }} 
\newcommand{\R}{{\mathbb R}}

\newcommand{\C}{{\mathbb C}} 
\newcommand{\ap}{ \alpha\pi^{-1}}

\newcommand{\Da}{{D}^\ast} 
\newcommand{\Ea}{{E}^\ast} 
\newcommand{\F}{{\mathcal F}} 
\newcommand{\Em}{{\mathcal E}} 
\newcommand{\Q}{{\mathcal Q}} 
\newcommand{\Ll}{{\mathcal L}} 
\newcommand{\err}{{\mathcal Err}} 
 
\newcommand{\Hh}{{\mathcal H}} 
\newcommand{\pb}{{\bar \phi}} 
\newcommand{\eps}{\varepsilon} 
 
\newcommand{\aan}{a_{\lambda}} 
\newcommand{\ac}{a^{\ast}_{\lambda}} 
 
\newcommand{\ean}{\varepsilon_{\lambda}} 
\newcommand{\half}{\mbox{$\frac{1}{2}$}} 
 
\newcommand{\al}{{\alpha}} 
\newcommand{\pa}{{\parallel}} 
 
\newcommand{\so}{{\Sigma_\al}} 
\newcommand{\vs}{ \sigma} 
\newcommand{\spf}{{\Sigma_\al}} 
\newcommand{\Tpf}{T} 
\newcommand{\Eal}{{\mathbf{E}_\alpha}} 
\newcommand{\as}{\sqrt{\alpha}} 
 
\newcommand{\Ow}{{\mathcal O}} 
\newcommand{\ora}{{|0\rangle}}

\newcommand{\la}{\Lambda}

\newcommand{\const}{{\rm const.}} 
\newcommand{\mA}{{\mathcal A}} 
\newcommand{\ua}{\uparrow}

\newcommand{\pe}{\psi_{n+1}}

\newcommand{\bh}{{\bf H_\al}} 
\newcommand{\be}{\beta} 
 
\numberwithin{equation}{section}

\begin{document} 
 
\title[One non-relativistic Particle] 
{One non-relativistic Particle coupled to a photon field} 
\author[Christian Hainzl]{Christian Hainzl$^1$} 
\address{Mathematisches Institut, LMU M\"unchen, 
Theresienstrasse 39, 80333 Munich, Germany} 
\email{hainzl@mathematik.uni-muenchen.de} 
\date{\today}

\begin{abstract} 
We investigate the ground state energy of an electron 
coupled to a photon field. 
First, we regard the self-energy of a \lq\lq free\rq\rq \ electron, which we 
describe by the Pauli-Fierz 
Hamiltonian. We show that, in the case of small values of the coupling 
constant $\alpha$, the leading order term 
is represented by  $2\ap  (\Lambda - \ln[1 + \Lambda])$. 
 
Next we put the electron in the field of an arbitrary external 
potential $V$, such that the corresponding Schr\"odinger operator 
$p^2 + V$ has at least one eigenvalue, and show that by coupling 
to the radiation field the binding energy increases, at least for 
small enough values of the coupling constant $\alpha$. 
Moreover, we provide concrete numbers for $\al$, the ultraviolet cut-off 
$\la$, and the radiative correction for which our procedure works. 

\end{abstract} 
\maketitle 
 
\footnotetext[1]{Marie Curie Fellow} 
 
\section{Introduction and main results} 
 
In recent times great effort was made to start a program 
with the goal to put Quantum Electrodynamics (QED) on a firm mathematical 
footing. Bach, Fr\"ohlich, and Sigal were 
the first who proved (in \cite{BFS1,BFS2,BFS3}), at least for small values 
of various parameters, 
that atoms and molecules have a ground state in the presence of a quantized 
radiation field. 
In the one particle case Griesemer, Lieb, and Loss in \cite{GLL} succeeded in 
removing 
the restrictions on the parameters. We refer to \cite{GLL} for an extensive 
list of references concerning 
this subject. 
 
In the present paper we are especially interested in the behavior of one 
electron
coupled to a photon field. We consider the \lq\lq free\rq\rq \ case as well 
as the presence of 
an arbitrary external potential $V$ with at least one negative eigenvalue in the 
electron case. 
We are interested in the {\it ground state energy} of such an electron when the coupling constant $\al$ 
is small and the ultraviolet cutoff parameter $\la$ is  fixed. 
 
Without radiation field the self-energy of an electron is simply given 
by the bottom of the spectrum of $ - \Delta$, which is equal to $0$. In the 
case of coupling to a photon field 
the situation changes dramatically and the self-energy is a complicated 
function depending on $\al$ and $\la$. The fact that evaluating this 
quantity is a highly non-trivial problem was pointed out in \cite{LL} by 
Lieb and Loss, who, 
in contrary to us, considered the case of fixed $\al$ and large values of 
the cutoff parameter $\la$. 
 
The paper is organized as follows. In the present section we state our main 
results 
on the self-energy and on the binding energy of an electron, which will be 
proved in Sections 3 and 4. In Section 2 we 
comment on our  notation and in Section 5 we evaluate the radiative 
correction in the case of a nucleus. We provide concrete numbers for 
$\al$ and $\la$ for which our procedure works. 
 
\subsection{Self-energy of an electron} 
The self-energy of an electron  is described by the Pauli-Fierz operator 
\begin{equation}\label{rpf} 
\Tpf = (p + \sqrt{\al}A(x))^2 + \as\vs B(x) + H_f. 
\end{equation}

We fix units such that  $\hbar=c=1$ and the electron mass $m=\half$. 
The electron charge is then given by 
$e=\as$, with $\alpha\approx 1/137$ the fine structure constant. 
In the present paper $\al$ plays the role of a small, dimensionless number 
which measures the coupling to the radiation field. 
Our results hold for 
{\it sufficiently small values } of $\al$. 
$\vs$ is the vector of Pauli matrices $(\sigma_1,\sigma_2,\sigma_3)$. 
Recall, the $\sigma_i$ are hermitian $2\times 2$  matrices 
and fulfill the anti-commutation relations $\sigma_i\sigma_j + \sigma_j\sigma_i 
= 
2I_{\C^2} \delta_{i,j}$. 
The operator $p= -i\nabla$ is the electron momentum 
while $A$ is the magnetic vector potential. The magnetic field is $B = {\rm 
curl} \ A$. 
 
The underlying Hilbert space is 
\begin{equation} 
\Hh = 
\Ll^{2}({\mathbb {R}}^{3};\C^2)\otimes \F_b(\Ll^2(\R^3;\C^2)) 
\end{equation} 
where 
$\F_b$ is the bosonic Fock space for the photon field. 
 
The vector potential is 
\begin{equation} 
A(x) = \sum_{\lambda = 1,2}\frac 1{2\pi} \int_{\R^3} 
\frac{\chi(|k|)}{|k|^{1/2}} \ean(k) \left[\aan(k) e^{ikx} + \ac(k) 
e^{-ikx}\right] dk, 
\end{equation} 
and  the corresponding magnetic field reads 
\begin{equation} 
B(x) = \sum_{\lambda = 1,2}\frac 1{2\pi} \int_{\R^3} 
\frac{\chi(|k|)}{|k|^{1/2}} (k\wedge i\ean(k)) \left[\aan(k) 
e^{ikx} - \ac(k) e^{-ikx}\right] dk, 
\end{equation} 
where the operators $\aan$ and  $\ac$ satisfy the usual commutation relations 
\begin{equation} 
[a_\nu (k),\ac(q)] = \delta(k-q)\delta_{\lambda,\nu}, \quad [\aan(k), 
a_\nu(q)] = 0, \ \ 
{\rm etc}. 
\end{equation} 
The vectors $\ean(k) \in \R^3$ are any two possible orthonormal 
polarization vectors perpendicular to $k$ which we assume 
to be either odd or even, i.e.
\begin{equation}\label{epscond} 
\eps_\lambda(-k) = \pm \eps_\lambda (k).
\end{equation}

The function $\chi(|k|)$ describes the ultraviolet cutoff on the 
wavenumbers $k$. We choose  $\chi$  to be the Heaviside 
function $\Theta(\Lambda -|k|/l_C)$, where $l_C= \hbar/(mc)$  is the Compton 
wavelength. Therefore $\la=1/4$ corresponds to the photon  energy $mc^2$, 
which sometimes is considered as a natural upper bound for the cut-off parameter, 
since it is the maximum value that guarantees that no pair-production takes 
place. 
 
The photon field energy $H_f$ is given by 
\begin{equation*} 
H_f = \sum_{\lambda= 1,2} \int_{\R^3} |k| \ac (k) \aan (k) dk. 
\end{equation*} 
 
We estimate the leading order in $\alpha$ of the ground state 
\begin{equation*} 
\spf = \infspec \Tpf. 
\end{equation*} 
Recall that $\al\langle 0 | A^2 |0 
\rangle = \ap \la^2$, where $|0 \rangle$ is the vacuum 
in $\F_b$, describes the vacuum fluctuation of the field $A$. Thus, this term 
is somehow ab initio included in the self-energy. Therefore 
it may turn out as a slight surprise that 
$ \ap  \la^2$ is not the leading order term of $\spf$, which relies on 
the fact that the magnetic field $\as \vs B$ reproduces 
the same number but with a negative sign. We show that the leading order 
term is given by $2\ap[\la - \ln(1+ \la)]$, 
which is the content of our first theorem. 
 
\begin{thm}\label{pf} 
Let $\la$ be fixed but arbitrary. Then 
\begin{equation} 
\left| \spf - 2\ap[\la - \ln(1+\la)]\right| \leq C_1 \al^2(\la^2+ \la^3 ) +C_2
\al^3 (\la^3 + \la^4)\ln(1+\la)
\end{equation} 
for some constants $C_{1,2} >0$ independent of $\al$ and $\la$. 
\end{thm}

It turns out in the proof that {\it one photon} is enough to 
recover the leading order term in $\al$. 
 
\subsection{Increase of the  binding energy} 
We consider an electron in an arbitrary external potential, 
described by a real valued function  $V$, independent of $\al$, which satisfies  (i) that 
 the negative part of $V$, $V_-$,  is dominated by the kinetic energy  $p^2 = 
-\Delta$, 
\begin{equation} 
V_- \leq -\eps \Delta + C_\eps, 
\end{equation} 
for any $\eps >0$ and for some positive constant $C_\eps$, 
such that the self-adjointness and boundedness from below of $p^2 + V$ is guaranteed. 
Additionally, since we want the essential spectrum to start at $0$, we require (ii) that 
$V_-$ tends to $0$ at infinity. 
Furthermore, we assume (iii) that 
the operator $p^2 +V$ has at least 
one negative energy bound state. 
 
The corresponding operator with a quantized radiation field reads 
\begin{equation} 
{\bf H_\al}= T + V, 
\end{equation} 
for which \cite{Hi} guarantees self-adjointness. 
Recently, Griesemer, Lieb, and Loss \cite{GLL} proved for all 
values of $\al, \la$ that $\bh$ has an eigenstate, as well as that 
the binding energy $ \spf - \Eal$, with 
\begin{equation} 
\Eal = \infspec \bh, 
\end{equation} 
cannot decrease, i.e. 
\begin{equation}\label{bc} 
\Eal - \spf\leq -e_0, 
\end{equation} 
where $-e_0 = \infspec [p^2 + V]$. We show that 
at least for small values of the coupling constant $\al$ the 
binding energy increases, namely (\ref{bc}) holds with strict 
inequality. 
 
\begin{thm}\label{t2} 
Let $V$ satisfy the conditions (i)-(iii), and 
$\phi$ be the ground state of $p^2 + V$ with corresponding energy $-e_0$. Then 
\begin{equation}\label{es} 
\Eal - \spf \leq -e_0 - \al \Em(V,\la) + \Ow(\al^2) 
\end{equation} 
for some positive number  $\Em(V,\la)$ depending on $ V$ and the cut-off $\la$. 
Moreover,  there exists a number $\rho > 0$, such that for all $\al 
\in (0,\rho]$ 
\begin{equation} 
\Eal - \spf < -e_0. 
\end{equation} 
\end{thm} 
\begin{rem} 
The equation (\ref{es}) holds for all real $V$, such that  $p^2 + V$ is 
self-adjoint and has  a ground state. 
But only for $V$ satisfying $(i) -(iii)$ we know that $H_\al$ has a ground state. 
 
We actually check  that  $\Em(V,\la)$ is bounded from above by 
\begin{equation} 
\Em(V,\la) \leq \pa p \phi\pa^2 \frac{32\pi}3 \ln[1+\la]. 
\end{equation} 
\end{rem} 
 
In the spinless case ($B=0$) it was previously proven
(\cite{HVV}) that coupling to a photon field can produce a ground state
even if the corresponding Schr\"odinger operator has only
continuous spectrum. This phenomenon of enhanced binding was earlier shown in
\cite{HS} in the case of dipole approximation ($A=A(0)$) in the limit of large
values of the coupling constant $\alpha$.

\section{Notation} 
 
Throughout the paper we use the notation 
\begin{equation} 
A(x)= D(x) + D^*(x), \, \, B(x)=E(x) + E^*(x) 
\end{equation} 
for the vector potential, respectively the magnetic field. 
 
The operators $D^*$ and $E^*$ create a photon with wave functions $G(k)e^{-i 
k\cdot x} $ and $H(k)e^{-ik\cdot x}$, respectively, where  $G(k)=(G^1(k),G^2(k))$ 
and $H(k)=(H^1(k),H^2(k))$ are vectors of one-photon states, given by 
\begin{equation}\label{defG} 
G^{\lambda}(k)=\frac {\chi(|k|)}{2\pi |k|^{1/2}} \eps_\lambda(k)  . 
\end{equation} 
and 
\begin{equation}\label{defH} 
H^{\lambda}(k)=\frac {-i\chi(|k|)}{2\pi |k|^{1/2}} k\wedge \eps_\lambda(k). 
\end{equation} 
Applying Fourier transform, with respect to $k$, on $G(k)e^{-i 
k\cdot x} $, i.e. 
\begin{equation} 
\F^{-1}[G(k)e^{-ik\cdot x}](\xi)= \int_{\R^3}e^{ik\cdot(\xi - x)}G(k) dk \equiv 
D^*|0\rangle (\xi - x), 
\end{equation} 
we see that  $D^*$ creates a photon depending on  $\xi- x$, where 
$\xi$ denotes the photon variable in configuration space. 
 
If the reader is not used in thinking of a \lq\lq position\rq\rq \ of a  photon,  he may consider it as a simple 
unitary transform. Because, now we introduce the variable 
$\eta= \xi - x$ and the object we are going to work with is the 
Fourier transform with respect to $\eta$, namely 
\begin{equation} 
\F[ D^*|0\rangle (\eta)](k) = G(k). 
\end{equation} 
The same result holds for $E^*$ if $G(k)$ is replaced by $H(k)$.

Notice that in the so chosen notation the electron momentum 
$p=-i\nabla_x$ of a one photon state $D^*f(x,\eta)= f(x)\otimes D^*|0\rangle (\eta)$, 
with $\eta=\xi-x$, can be written as 
\begin{multline} 
p D^*f(x,\eta) = \sum_{i=1}^3\big(p^i_x f(x)\big)\otimes 
D_i^*|0\rangle (\eta) - f(x)\otimes\big(p_\eta D^*|0\rangle 
(\eta)\big) \\= [p_x - p_\eta] D^*f(x,\eta). 
\end{multline}

It turns out to be convenient to denote a general vector  $\Psi 
\in \Hh$ as a sequence 
\begin{equation} 
\Psi= \{\psi_0, \psi_1, \psi_2,...\}, 
\end{equation} 
where 
\begin{equation} 
\psi_n = \psi_n(x,\xi_1-x,...,\xi_n-x), 
\end{equation} 
with relative variables $\xi_i -x=\eta_i$. 
For simplicity, we suppress the polarization of the photons and the spin of 
the 
electron.

\section{Proof of Theorem \ref{pf} } 
\subsection{Upper bound} 
We choose the sequence of trial wave functions 
\begin{equation}\label{tf} 
\Psi_n = \{f_n(x)  \ua, -\as  [p_{\eta}^2 + H_f]^{-1}(\vs \ua) \Ea 
f_n(x,\eta) , 0,0,\ldots \}, 
\end{equation} 
with $\eta=\xi-x$, $f_n \in \Ll^2(\R^3)$, $\| f_n\|=1$ and $\ua$ denoting the spin-up vector 
in $\C^2$. We assume for the electron part $f_n(x)$ that $\| pf_n\| \to 0$ as $n\to\infty$, 
which means that the electron spreads out to infinity. 
This is 
reasonable, for in the case without coupling this sequence recovers the 
bottom of the spectrum  of $-\Delta$. 
Recall, 
\begin{equation} 
\Ea  f_n(x,\eta) \equiv f_n(x) 
\otimes [\Ea |0\rangle] (\eta) 
\end{equation} 
and, due to $\eta=\xi -x$, 
\begin{equation} 
\left([p^2 + H_f]\Ea f_n\right)(x,\eta) = ([p_x - p_\eta]^2 + |p_\eta|)\Ea f_n(x,\eta). 
\end{equation} 
Notice, in the one photon case $H_f = |p_\eta|$. Observe, 
\begin{multline} 
[p,D]= \sum_{\lambda=1,2} \int_{\R^3 } G^\lambda(k)\cdot 
[p,e^{ik\cdot x}]a_\lambda(k) dk= \\ =  \sum_{\lambda=1,2} 
\int_{\R^3 } (G^\lambda(k)\cdot k) e^{ik\cdot x}a_\lambda(k) dk=0 
\end{multline} 
by our choice of $\eps_\lambda(k)$. Analogously, 
\begin{equation}\label{compd} 
[p,D^*]=[p,E]=[p,E^*]=0. 
\end{equation} 
By means of Schwarz' inequality it is easy to see that by our 
assumption $\|pf_n\|\to 0$ 
\begin{equation} 
\lim_{n\to\infty} (\Psi_n,p  A\Psi_n) = 2\lim_{n\to\infty}\Re (\Psi_n,p D\Psi_n) =0. 
\end{equation} 
Similarly, if we denote the $1$-photon part of $\Psi$ as 
\begin{equation} 
\psi^n_1 = -\as [p_\eta^2 + H_f]^{-1}(\vs \ua) \Ea f_n, 
\end{equation} 
then 
\begin{equation} 
(\psi^n_1 ,p^2 \psi^n_1)  = (\psi^n_1,[p_x+p_\eta]^2 \psi_1^n)= 
(\psi^n_1 ,p_\eta^2 \psi^n_1) + \mbox{Error}(p_x f_n), 
\end{equation} 
with $\lim_{n\to\infty} \mbox{Error}(p_x f_n)= 0$. Furthermore, 
we compute 
\begin{multline}\label{zizu} 
\big(\psi_1^n,[p_\eta^2 + H_f]\psi_1^n\big) + 2\as \Re \big( 
(\vs \ua) \Ea f_n, \psi_1^n \big) \\= -\al \big( (\vs \ua) \Ea f_n, 
[p_\eta^2 + H_f]^{-1} (\vs \ua)  \Ea f_n\big)\\ = - \al\langle 0| 
(\vs \ua) E [p^2 + H_f]^{-1} (\vs \ua)  \Ea\ora 
\end{multline} 
In momentum representation, recall 
 $\F[E^*\ora](k) = H(k)$, 
the last term in (\ref{zizu}) equals 
\begin{multline} 
\langle 0| (\vs \ua) E [p^2 + H_f]^{-1} (\vs \ua)  \Ea\ora=\int 
\frac{|H(k)\cdot(\vs \ua)|_{\C^2}^2}{|k|^2+|k|} dk \\=\int 
\frac{|H(k)|^2}{|k|^2+|k|} dk = \langle 0|E [p^2 + 
H_f]^{-1}\Ea\ora . 
\end{multline} 
The equality from first to second line holds, because of the 
anti-commutation relations of $\vs$ and the fact that $H$ is 
purely imaginary. 
 
For simplicity, we will use throughout the paper 
\begin{equation} 
|H(k)|^2= \sum_{\lambda=1,2} |H^\lambda(k)|^2, \ \ |G(k)|^2= \sum_{\lambda=1,2} |G^\lambda(k)|^2. 
\end{equation} 
 We evaluate 
\begin{multline} 
\langle 0|E [p^2 + H_f]^{-1}\Ea\ora =\sum_{\lambda =1,2} \int 
\frac{| H^\lambda(k)|^2}{|k|^2 + |k|} dk \\ = \frac 1{2\pi^2} \int 
\frac{|k|\chi(|k|)}{|k|^2 + |k|} dk = \pi^{-1}[\la^2 - 2[\la - 
\ln(1 + \la)]]. 
\end{multline} 
Since 
\begin{equation}\label{aqq} 
[D,D^*] \equiv D D^* - D^*  D = \pi^{-1}\la^2, 
\end{equation} 
we know 
\begin{equation} 
A^2=(D+D^*)^2= \pi^{-1} \Lambda^2 + 2 D^* D +D^* D^* + D D. 
\end{equation} 
With $(\Psi_n,DD\Psi_n) =0$ and $\lim_{n\to \infty} (\psi_1^n,\psi_1^n) \leq 2\al 
\pi^{-1} \ln(1+\la)$
we arrive at 
\begin{multline} 
 \lim_{n\to \infty} (\Psi_n,T\Psi_n) \leq \ap \la^2 - \al 
\langle 0|E [p^2 + H_f]^{-1}\Ea\ora + 2\pi^{-2} \al^2 \la^2 \ln(1+\la)\\ +2 \al^2  \pa D [p^2 + 
H_f]^{-1} (\vs \ua)  \Ea \ora\pa^2. 
\end{multline} 
The last term in the r.h.s. vanishes which can be seen by explicit calculations
using the relation
\begin{equation}
\sum_{\lambda=1,2} \eps_{\lambda}^i  \eps_{\lambda}^j = \delta_{i,j} - \frac{k_ik_j}{|k|^2}.
\end{equation} 
Namely, if $\epsilon^{j \, l\, n}$ denotes the totally antisymmetric 
epsilon-tensor,
we obtain
\begin{multline}\label{epstens}
\sum_{\lambda =1,2} \int \frac{G^{\lambda}_i (k) H^\lambda_j (k)}{|k|^2 + |k|}dk
= \sum_{\lambda =1,2} \sum_{l,n=1}^{3}i\int \frac{\chi(|k|) \eps_\lambda^i(k)\big[
\epsilon^{j \, l\, n} \eps_\lambda^l(k) k_n\big]}{|k|^3 + |k|^2}dk\\
=  \sum_{l,n=1}^{3}i\int \frac{\chi(|k|)\big[\delta_{i,l} - \frac{k_ik_l}{|k|^2}\big]
\epsilon^{j \, l\, n}  k_n}{|k|^3 + |k|^2}dk = 0.
\end{multline}
Therefore, since $(\Psi_n,\Psi_n) = 1 + \const \al$, we obtain the upper bound 
\begin{multline} 
\spf \leq 
 \lim_{n\to \infty} (\Psi_n,T\Psi_n)/(\Psi_n,\Psi_n) \leq  \lim_{n\to \infty} 
(\Psi_n,T\Psi_n)\leq \\ 
\leq 2\ap  [\la - \ln(1+\la)] +  2\pi^{-2} \al^2 \la^2 \ln(1+\la). 
\end{multline}

\subsection{Lower bound} 
We have learned in the previous section that an approximate ground state $\Psi$, 
$\|\Psi\| =1$, satisfies 

\begin{equation}
(\Psi, \Tpf \Psi) \leq 2\ap\la +2\pi^{-2} \al^2 \la^2 \ln(1+\la) .
\end{equation}
 Since 
\begin{equation} 
\Tpf= [ \vs\cdot (p + \as A(x))]^2 + H_f \geq H_f 
\end{equation} 
we immediately get 
\begin{equation}\label{aprih} 
(\Psi, H_f\Psi) \leq 2\ap \Lambda +2\pi^{-2} \al^2 \la^2 \ln(1+\la) . 
\end{equation} 
Therefore, by means of Schwarz' inequality we 
obtain 
\begin{eqnarray}\label{39} 2\as(\Psi,p A\Psi)&= &4 \as 
\Re (\Psi,p D \Psi) \leq a\pa p\Psi\pa^2 + 4a^{-1}\al \pa 
D\Psi\pa^2\\ \nonumber \as(\Psi,\vs  B\Psi)&=& 2\as(\Psi,\vs 
 E\Psi) \leq c \la^2 \al \pa \Psi\pa^2 + c^{-1} (1/\la^2) \pa 
E \Psi\pa^2,\\ \label{310} 
\end{eqnarray} 
for any $a,c > 0$. 
Since, by $A^2\geq 0$, 
\begin{multline} 
\|p\Psi\|^2 \leq (\Psi,\Tpf \Psi) - (\Psi,H_f\Psi) - 4 \as \Re 
(\Psi,p D \Psi) - 2\as(\Psi,\vs   E\Psi), 
\end{multline} 
we obtain by (\ref{39}), (\ref{310}),  the operator inequalities 
(e.g. \cite[Lemma A. 4]{GLL}) 
\begin{equation}\label{energy} 
\Da D \leq \frac 2\pi \la H_f, \quad \Ea E \leq \frac 2{3\pi} 
\la^3 H_f, 
\end{equation} 
and (\ref{aprih}) 
\begin{equation}\label{aprip} 
\pa p\Psi\pa^2 \leq C_1\al (\la+\la^2) +C_2\al^2 (\la^2 + \la^3)\ln(1+\la)
\end{equation} 
for some suitable constants $C_{1,2}>0$. Equations (\ref{aprih}) 
and (\ref{aprip}) will be decisive to control our error estimates. 
Recall, 
\begin{equation} 
A^2 = \pi^{-1}\la^2 + 2 \Da D +  2 \Re D D. 
\end{equation} 
Thus, since $\Da D \geq 0$, 
\begin{equation} 
(\Psi,T\Psi) \geq \ap \Lambda^2 \pa \Psi \pa^2 + 
\Em_0[\psi_0,\psi_1] + \sum_{n =0}^{\infty} 
\Em[\psi_n,\psi_{n+1},\psi_{n+2}]  , 
\end{equation} 
where 
\begin{equation} 
\Em_0 [\psi_0,\psi_1] = ( \psi_1, \mA \psi_1)  + 
2\as{\Re}\big ([\vs \Ea + 2pD^*]\psi_0,\psi_1\big) 
\end{equation} 
and 
\begin{multline} 
\Em[\psi_n,\psi_{n+1},\psi_{n+2}] = (\psi_{n+2} , \mA \psi_{n+2}) 
\\ + 2\Re\left(\Big([\as \vs  \Ea + 2\as p \Da]\psi_{n+1} 
+ \al \Da\Da \psi_n\Big), \psi_{n+2}\right), 
\end{multline} 
with 
\begin{equation} 
\mA= p^2 + H_f. 
\end{equation} 
 
Recall, we can write our approximate ground state $\Psi$ as 
\begin{equation} 
\Psi = \{\psi_0 (x), \psi_1(x,\eta_1),....,\psi_n(x,\eta_1,...,\eta_n),...\}, 
\end{equation} 
with $\eta_i=\xi_i -x$. 
For convenience we will work in momentum representation 
\begin{equation} 
\F[\psi_n(x,\eta_1,...,\eta_n)](l,k_1,...,k_n) = 
\psi_n(l,k_1,\dots ,k_n), 
\end{equation} 
with $k = (k_1,...,k_n)$ and $l$ denotes the momentum representation 
of the electron variable.

We consider the term $\Em[\psi_n,\psi_{n+1},\psi_{n+2}]$. As a 
straightforward consequence of Schwarz' inequality we derive 
\begin{multline}\label{schw} 
\Em[\psi_n, \psi_{n+1}, \psi_{n+2}]\geq\\ \geq- \Big\| \as 
\big[\mA^{-1/2} \vs \Ea + \mA^{-1/2}p \Da \big ]\psi_{n+1} +\al 
\mA^{-1/2} \Da \Da \psi_n \Big\|^2. 
\end{multline} 
Similarly, 
\begin{equation}\label{eo} 
\Em_0[\psi_0,\psi_1] \geq -\al\big\|( \mA^{-1/2}\vs \Ea + 2 \mA^{-1/2}pD^*) \psi_0\big\|^2 
\end{equation} 
The fact that $H(k)$ is purely imaginary and $\mA$ commutes with the reflection 
$l\to -l$ 
imply 
\begin{equation} 
\Im(E_i^*\psi_0,\mA^{-1}E_j^*\psi_0) = 0. 
\end{equation} 
Together with the anti-commutation relations for $\vs$ we infer 
\begin{equation} 
 (\vs \Ea \psi_0, \mA^{-1} \vs \Ea \psi_0)=(\psi_0, E\mA^{-1} \Ea\psi_0). 
\end{equation}

Now we are going to evaluate the r.h.s. of (\ref{schw}). These evaluations then are also applied to (\ref{eo}). 
First, we consider the diagonal terms. 
The most important one, that reproduces the leading order in $\al$,  is 
\begin{equation}\label{imp} 
-\al (\vs \Ea \pe, \mA^{-1} \vs \Ea \pe). 
\end{equation} 
 
Explicitly, in Fourier representation, 
\begin{equation} 
\F[E^*\psi_{n+1}](l,k_1,...,k_{n+2})= \frac 
1{\sqrt{n+2}}\sum_{i=1}^{n+2}H(k_i)\psi_{n+1} 
(l,k_1,\dots,\not \!\! k_i,\dots,k_{n+2}), 
\end{equation} 
where $\not \!\! k_i$ indicates that the $i$-th variable is omitted in 
$\psi_{n+1}$. Due to permutational symmetry the expression (\ref{imp}) consists of 
two different terms 
\begin{equation}\label{pablo} 
-\al (\vs \Ea \pe, \mA^{-1} \vs \Ea \pe)= -\alpha \big(I_{n+1} + II_{n+1}) . 
\end{equation} 
The diagonal term $I_{n+1}$ where in the right as well as left hand side of 
(\ref{imp}) a photon is created with index $i$ and the mixed term 
$II_{n+1}$. By similar arguments as for (\ref{eo}), namely if we take advantage of 
the fact that $H$ is purely imaginary and use the anti-commutation relations of 
$\vs$ we obtain 
\begin{equation}\label{laprochita} 
I_{n+1} \leq \int 
\frac{|H(k_{n+2})|^2|\psi_{n+1}(l,k_1,\dots,k_{n+1})|^2}{[l-\sum_{i=1}^{n+2} 
k_i]^2 +\sum_{i=1}^{n+2} |k_i|} 
dldk_1\dots dk_{n+2}. 
\end{equation} 
We set $\Q = [l-\sum_{i=1}^{n+1} 
k_i]^2+ |k_{n+2}|^2 +\sum_{i=1}^{n+2} |k_i|$ and $b= 2 [l-\sum_{i=1}^{n+1} 
k_i]\cdot k_{n+2}$ and expand 
\begin{equation} 
\frac 1{\Q- b} = \frac 1{\Q} + \frac 1{\Q}b\frac 1{\Q} +  \frac 1{\Q}b\frac 1{\Q-b}b\frac 1{\Q}. 
\end{equation} 
Plugging this expansion into (\ref{laprochita}), the second term, when 
integrating over $k_{n+2}$, obviously vanishes. Since $\Q \geq |k_{n+2} |^2 
+|k_{n+2} |$ and $\Q -b  \geq |k_{n+2} |$ we estimate 
\begin{multline}\label{61} 
I_{n+1} \leq \|\psi_{n+1}\|^2 \int_{\R^3} \frac{|H(k)|^2}{|k|^2 + 
|k|}dk \\ + 4  \int \frac{|H(k_{n+2})|^2|k_{n+2}|^2 
|l-\sum_{i=1}^{n+1} k_i|^2 |\psi_{n+1}(l,k_1,\dots,k_{n+1})|^2}{ 
[|k_{n+2} |^2 +|k_{n+2} |]^2 |k_{n+2} |}\\ \leq  \|\psi_{n+1}\|^2 
\langle 0|E \mA^{-1} E^* \ora + \const \la \| p \psi_{n+1}\|^2. 
\end{multline} 
 
For the mixed term we have 
\begin{multline}\label{II} 
II_{n+1} \leq 3 (n+1) \sum_{\lambda=1,2} \int \frac{ 
|H^{\lambda}(k_1)||H^{\lambda}(k_{n+2})|}{[l-\sum_{i=1}^{n+2} 
k_i]^2 +\sum_{i=1}^{n+2} |k_i|}\times \\ \times 
|\psi_{n+1}(l,k_2,\dots,k_{n+2})| 
|\psi_{n+1}(l,k_1,\dots,k_{n+1})| dldk_1\dots dk_{n+2}. 
\end{multline} 
By means of the one-photon density 
\begin{equation}\label{opd} 
\rho_{\pe}(k) = (n+1) \int 
|\psi_{n+1}(l,k,k_2,\dots,k_{n+1})|^2dldk_2\dots dk_{n+1} 
\end{equation} 
we infer after applying Schwarz' inequality to (\ref{II}) 
\begin{multline}\label{64} 
II_{n+1} \leq 3 (n+1) \sum_{\lambda=1,2} \int \frac{ 
|H^{\lambda}(k_1)||H^{\lambda}(k_{n+2})|}{|k_1| + |k_{n+2}|}\times 
\\ \times 
\sqrt{\rho_{\pe}(k_1)}\sqrt{\rho_{\pe}(k_{n+2})}dk_1 dk_{n+2} 
\\ \leq  \frac { 6(n+1) }{(2\pi)^2}\int \frac{\chi(|k_1|)\chi(|k_{n+2}|)}{|k_1| + |k_{n+2}|} 
\sqrt{|k_1|\rho_{\pe}(k_1)}\times \\ \times \sqrt{|k_{n+2}|\rho_{\pe} (k_{n+2})}dk_1dk_{n+2} 
\\ \leq \frac 3{2\pi^2} \int |k| \rho_{\pe}(k) dk \left[ \int 
\frac{\chi(|k_1|)\chi(|k_{n+2}|)}{(|k_1| + 
|k_{n+2}|)^2}dk_1dk_{n+2}\right]^{1/2} 
\\ \leq \const \la^2 (\psi_{n+1} ,H_f\psi_{n+1}). 
\end{multline} 
Therefore, we summarize 
\begin{multline}\label{estar} 
-\al (\vs \Ea \pe, \mA^{-1} \vs \Ea \pe)\geq -\al \langle 0|E 
 \mA^{-1} E^*\ora\|\pe\|^2 \\ -  \const \al \la \Big(\|p\psi_{n+1}\| +\la 
 (\psi_{n+1}, H_f \psi_{n+1})\Big). 
\end{multline} 
 
All other terms in (\ref{schw}) are of order $\Ow(\al^2)$ or even of higher order. 
 
For the second diagonal term we obtain 
\begin{equation}\label{diego} 
-\al^2(\psi_n, D D \mA^{-1} \Da \Da \psi_n)\geq -\const \al^2\la 
\Big(\la\|\psi_n\|^2 + \|p\psi_n\|^2 + (\psi_n,H_f\psi_n)\Big). 
\end{equation} 
The proof of (\ref{diego}), whose strategy is similar to the one 
for (\ref{estar}), will be postponed to Lemma \ref{DD} in the appendix. 
 
The third diagonal term of (\ref{schw}) reads 
\begin{multline}\label{mc} 
 -\al (p\Da \pe, \mA^{-1} p \Da \pe) = 
\sum_{\lambda =1,2} -\al \times \\ \times \Big[ \int \frac{ 
\left[G^\lambda(k_{n+2}) \cdot \big(l- \sum_{i=1}^{n+1} k_i\big ) 
\right]^2 |\pe(l,k_1,\dots,k_{n+1})|^2}{\left|l- \sum_{i=1}^{n+2} 
k_i\right|^2  + \sum_{i=1}^{n+2} |k_i|}dldk_1\dots dk_{n+2}  \\ + 
(n+1) \int \frac{ \left[G^\lambda(k_1) \cdot \left(l- 
\sum_{i=1}^{n+2} k_i\right) \right] \left[G^\lambda(k_{n+2}) \cdot 
\left(l- \sum_{i=1}^{n+2} k_i\right)\right]}{\left|l- 
\sum_{i=1}^{n+2} k_i\right|^2 +\sum_{i=1}^{n+2} |k_i|} \times \\ 
\times \overline{\pe(l,k_1,\dots,k_{n+1})}\pe(l,k_2,\dots,k_{n+2}) 
dldk_1\dots dk_{n+2}\Big] \\ \geq -\const \al\la\Big(\|p\pe\|^2 
+(\pe,H_f\pe)\Big). 
\end{multline} 
For the second term in the r.h.s. we used first 
\begin{equation} 
\frac{\left|l- \sum_{i=1}^{n+2} k_i\right|^2}{\left|l- 
\sum_{i=1}^{n+2} k_i\right|^2 +\sum_{i=1}^{n+2} |k_i|} \leq 1, 
\end{equation} 
and afterwards applied Schwarz' inequality to bound it from below by 
\begin{multline}\label{ddhf} 
\geq-(n+1)\sum_{\lambda=1,2} \int \left|G^\lambda(k_1) \right| \left|G^\lambda(k_{n+2})\right|\sqrt{ \rho_{\pe}(k_1)\rho_{\pe}(k_{n+2})} 
dk_1 dk_{n+2} \\ \geq-  \const \int |k| \rho_{\pe}(k) dk \left[ \int 
\frac{\chi(|k_1|)\chi(|k_{n+2}|)}{|k_1|^2|k_{n+2}|^2}dk_1dk_{n+2}\right]^{1/2}, 
\end{multline} 
which yields the last term in the last line of (\ref{mc}). 
 
Next we consider the off-diagonal terms. Notice, 
\begin{multline}\label{hugoo} 
\al\Re (p\Da \pe, \mA^{-1}\vs \Ea \pe) =\\ 
=\sum_{\lambda=1,2}\al\Big[ \Re \int \frac{ 
\left[G^\lambda(k_{n+2}) \cdot \big(l- \sum_{i=1}^{n+1} k_i\big 
)\right] H^\lambda(k_{n+2})\cdot }{\left| l- \sum_{i=1}^{n+2} 
k_i\right|^2+ \sum_{i=1}^{n+2} |k_i|} \times \\ \times \cdot \langle 
\pe,\vs \pe\rangle_{\C^2}(l,k_1,\dots,k_{n+1}) dldk_1\dots 
dk_{n+2} \\ + (n+1) \Re \int \frac{ \left[G^\lambda(k_1) \cdot 
\big(l- \sum_{i=2}^{n+2} k_i\big ) \right] 
H^\lambda(k_{n+2})\cdot} { \left | l- \sum_{i=1}^{n+2} k_i\right|^2 
+\sum_{i=1}^{n+2} |k_i|} \times \\ 
\times \cdot \langle \pe(l,k_2,\dots,k_{n+2}),\vs \pe(l,k_1,\dots,k_{n+1})\rangle_{\C^2} dldk_1\dots dk_{n+2}\Big]. 
\end{multline} 
The first term in the r.h.s. vanishes, because the integral is 
purely imaginary. For the second term we use $\frac {|a|}{a^2 + b} 
\leq \half b^{-1/2}$ and Schwarz' inequality to bound it from 
above by 
\begin{multline} 
(\ref{hugoo}) \leq \al \sum_{\lambda =1,2} \int 
\frac{|G^\lambda(k_1)||H^\lambda(k_{n+2})|} 
{(|k_1| + 
|k_{n+2}|)^{1/2} |k_1|^{1/2} |k_{n+2}|^{1/2}} \times\\ \times 
\sqrt{\rho_{\pe}(k_1) |k_1|} \sqrt{\rho_{\pe}(k_{n+2}) |k_{n+2}|} 
dk_1dk_{n+2} 
\\ \leq \const \al \la^{3/2} (\pe, H_f \pe). 
\end{multline} 
 
The second  off-diagonal term, $2\al^{3/2} \Re(p\Da \pe, \mA^{-1} \Da\Da 
\psi_n)$, can simply be bounded from above by 
\begin{equation}\label{72} 
\al (\pe, p D \mA^{-1}p\Da \psi_{n+1}) + \al^2 (\psi_n 
,DD\mA^{-1} \Da\Da \psi_n), 
\end{equation} 
on which we now apply (\ref{diego}) and (\ref{mc}). 
 
Finally, by means of Lemma \ref{off} in the appendix we estimate the last off-diagonal 
term by 
\begin{multline} 
\al^{3/2}\Re (\vs \Ea \pe, \mA^{-1} \Da\Da \psi_n)\leq \const \al 
\la \Big(\al \|\pe\|^2 +  (\psi_n, H_f \psi_n)\Big). 
\end{multline} 
Collecting above estimates and summing over all $n$ we arrive at 
\begin{multline} 
(\Psi, T\Psi) \geq 2\ap\big(1 - \const \al (1+\la)\big) \la^2 \pa\Psi\pa^2 - 
\al\pa \Psi\pa^2 \langle 0|E \mA^{-1} \Ea | 0 \rangle \\- \const 
\al\la(1+\la) (\Psi, H_f\Psi) - \const\al \la \pa p\Psi\pa^2 . 
\end{multline} 
By our a priori estimates (\ref{aprih}) and (\ref{aprip}) we prove 
the theorem. 
 
\section{Proof of Theorem  \ref{t2}} 
 
It suffices to use a cleverly chosen trial wave function. 
We assume that $\phi(x) \in \Ll^2(\R^3)$ is the 
ground state of $p^2 + V$, i.e. 
\begin{equation} 
(p^2 + V)\phi= -e_0 \phi. 
\end{equation} 
Recall, 
\begin{equation*} 
\Ea \phi(x, \eta) = \phi (x)\otimes [\Ea (x)|0\rangle](\eta),\quad 
\Da p\phi= \sum_{i=1}^3 p_x^i\phi (x) \otimes [\Da_i 
|0\rangle](\eta) 
\end{equation*} 
are 1-photon functions depending on $x$ and the relative 
coordinates $\xi-x = \eta$. Therefore, in 
configuration space, where $\psi(x,\eta)$ denotes one of these 
functions, we have (cf. the previous section) 
\begin{equation} 
[(p^2 + H_f)\psi](x,\eta)= ([p_x - p_\eta]^2 + 
|p_\eta|)\psi(x,\eta). 
\end{equation} 
For sake of convenience we define the operator 
\begin{equation} 
A_V =p_x^2 + p^2_\eta +H_f +V + e_0= (p_x^2 + V +e_0)\otimes 
\mathbb{I} + \mathbb{I}\otimes (p_\eta^2 + H_f), 
\end{equation} 
which acts on $\Ll^2\left((\R^3;\C^2)\otimes( \R^3;\C^2)\right) $ 
and is obviously positive and invertible. 
 
Now, we choose our trial wave function $\Psi \in \Hh$ as 
\begin{equation} 
\Psi = \{\pb(x), -2\as A_V^{-1}\Da p\pb(x,\eta) - \as A_V^{-1} \vs \Ea \pb(x,\eta), 
0, 0,...\}, 
\end{equation} 
where $\pb= \phi  \ua$ and $\eta=\xi-x$. We assume $\pa \phi\pa =1$ and for 
simplicity we denote the 1-photon part of $\Psi$  as $\psi_1$. Notice, \\$\vs 
\Ea \pb(x,\eta)= \phi(x) \otimes (\vs \ua)  \Ea |0\rangle(\eta)$. 
 
First, observe that $A_V \pb = 0$ yields 
\begin{equation} 
A_V \vs \Ea \pb=(p_\eta^2 + H_f)  \vs \Ea \pb. 
\end{equation} 
Therefore, since $A_V$ and $p_\eta^2 + H_f$ commute by definition, 
we infer 
\begin{equation} 
A_V^{-1}  \vs \Ea \pb= (p_\eta^2 + H_f)^{-1} \vs \Ea \pb, 
\end{equation} 
and 
\begin{equation} 
\left(\pb,  \vs E A_V^{-1}  \vs \Ea \pb\right) = \pa \phi\pa^2 
\langle 0|E [p_\eta^2 + H_f]^{-1}\Ea|0\rangle. 
\end{equation} 
Thus, we evaluate 
\begin{multline} 
(\Psi, \bh \Psi) =\ap \la^2 \pa \Psi\pa^2 - e_0 \pa \Psi\pa^2 + 
(\psi_1, A_V \psi_1)\\ + 2\as \Re \left((\vs \Ea + 2\Da p)\pb, 
\psi_1\right) + 2\al(\psi_1,\Da D \psi_1) + 2 (\psi_1,p_x \cdot 
p_\eta \psi_1). 
\end{multline} 
Notice, the cross term 
\begin{equation} 
\Re(\pb, \vs E A_V^{-1} \Da p \pb) = \Re(\pb, \vs E [p_\eta^2 + 
H_f]^{-1} \Da p\pb) 
\end{equation} 
vanishes as in the previous section, because it is purely 
imaginary. Therefore, by our choice of $\psi_1$ we get 
\begin{multline} 
(\psi_1, A_V \psi_1)+ 2\as \Re \left((\vs \Ea + 2\Da p)\pb, 
\psi_1\right)\\ = -4\al (\Da p_x \phi, A_V^{-1}\Da p_x \phi) - \al 
\pa\phi\pa^2 \langle 0| E [p_\eta^2 + H_f]^{-1}\Ea |0\rangle. 
\end{multline} 
Moreover, we have 
\begin{multline} 
(\psi_1,p_x \cdot p_\eta \psi_1) = \al (\Da p_x \phi, A_V^{-1} 
p_x\cdot p_\eta A_V^{-1} \Da p_x \phi) \\ - \al (\vs \Ea \pb, 
[p_\eta^2 + H_f]^{-1} p_x\cdot p_\eta [p_\eta^2 + H_f]^{-1}\vs \Ea 
\pb)\\ +2\al\Re (\Da p_x \pb, A_V^{-1}p_x\cdot p_\eta A_V^{-1}\vs 
\Ea \pb). 
\end{multline} 
The first term in the r.h.s. vanishes by integrating over the 
$\eta$-variable (the best way to see it is using the 
representation in momentum space), due to (\ref{epscond}) and 
the fact that the operator $A_V$ commutes 
with reflection $\eta \to -\eta$, respectively $k\to -k$. The 
second term vanishes when integrating over the $x$-variable 
(notice, $(\phi, p_x\phi)=0$), and the third term vanishes, 
because it is again purely imaginary. 
 
Furthermore, since $DA_V^{-1} \vs \Ea \pb =0$ (see (\ref{epstens})), we realize, after 
straightforward calculation, that 
\begin{equation} 
\al(\psi_1,D^*D\psi_1) \leq \const \al^2 \ln(1+\la)^2. 
\end{equation} 
Using above estimates and the fact that 
\begin{equation} 
\ap \la^2 - \al \langle 0 |E[p_\eta^2 + H_f]^{-1}\Ea|0\rangle 
= 2\ap [\la -\ln(1+\la)] 
\end{equation} 
we infer 
\begin{multline} 
(\Psi, \bh \Psi) /(\Psi,\Psi) \leq -e_0 + 2\ap [\la 
-\ln(1+\la)]\\ -4 \al (\Da p_x \phi, A_V^{-1}\Da p_x \phi)+ 
\Ow(\al^2)\ln(1+\la)(\la^2 + \ln(1+\la))
\end{multline} 
which proves the first statement of the theorem with 
\begin{equation}\label{rc} 
\Em(V,\la) = 4 \al (\Da p_x \phi, A_V^{-1}\Da p_x \phi). 
\end{equation}

The second statement follows by the observation that  $(\Da p_x 
\phi, A_V^{-1}\Da p_x \phi)$ is strictly larger than $0$, which is 
a consequence of the fact that $\Da p_x \phi $ is a not 
identically vanishing function $\in \Ll^2(\R^3 \otimes 
(\R^3,\C^2))$ and $A_V$ an invertible operator.

\section{Computation of concrete numbers}\label{shift} 
 
\subsection{Error for the self-energy} 
 
We are going to calculate the error $\err$ of the self-energy, 
\begin{equation} 
|\so - 2\ap [\la - \ln(1+\la)]| \leq \err(\al^2). 
\end{equation} 
First, it is important to estimate the kinetic energy term $\pa p\Psi\pa^2$ 
for an approximate ground state $\Psi$ with quite good constants. Since we will
calculate anyway with values $\la \leq 1$ we can assume $(\Psi,T\Psi) \leq 2\ap  \la$.
 
By means of (\ref{39}), (\ref{310}) and then applying (\ref{energy}) we obtain 
\begin{multline} 
2\ap  \la \geq (\Psi,T\Psi) \geq (1-a) \pa p\Psi\pa^2 + (\Psi,H_f\Psi) \\ 
- 4a^{-1}\al\pa D\Psi\pa^2 -c\al \la^3 - 1/(c\la^3) \pa E\Psi\pa^2 
\\ \geq (1-a)\pa p\Psi\pa^2 - c \al\la^3 +\left[ 1 - a^{-1} 8\pi^{-1} 
\al \la - c^{-1} \frac{2}{3\pi}\right](\Psi,H_f\Psi), 
\end{multline} 
where $ 1 > a > 0$. 
We require the last term $[..]$ to be $\geq 0$. For simplicity, we choose $c 
=\frac 2\pi$, then our first condition on $\al$ reads 
\begin{equation}\label{req} 
\al \leq \frac {a\pi}{4\la} 
\end{equation} 
and additionally 
\begin{equation}\label{ppsi} 
\pa p\Psi\pa^2 \leq \frac{2\al\la(1+\la^2)}{\pi(1-a)}. 
\end{equation} 
 
The main contribution to $\err$ stems from the 
 the terms  (\ref{estar}) and (\ref{mc}). 
In fact the the third diagonal term (\ref{diego}) is  negligible compared 
to (\ref{estar}) and (\ref{mc}). 
 
Evaluating the corresponding integrals yields 
\begin{equation} 
(\ref{mc}) \leq 2\ap \la \pa p\Psi\pa^2 + 2\ap \la (\Psi, H_f\Psi), 
\end{equation} 
and 
\begin{equation} 
 (\ref{61}) + (\ref{64}) \leq \frac 8{3\pi}\al \la \|p\Psi\|^2 + \frac 3{2\pi}\al \la^2  (\Psi, H_f\Psi). 
\end{equation} 
In order to incorporate the off-diagonal terms we, for simplicity, double the so gained value for  $\err$ 
and derive 
\begin{eqnarray} 
\err &\leq& 2\left[ \frac{14}{3\pi}\al  \la \pa p\Psi\pa^2 + \frac {7}{2\pi}\al \la^2 (\Psi, H_f\Psi)\right]\\ 
&\leq& \al^2 \left[ \frac{56\la^2(1+\la^2)}{3\pi^2(1-a)} +  14{\pi^2} \la^2\right] 
\end{eqnarray} 
where we used  (\ref{aprih}) and  (\ref{aprip}). 
 
\subsection{Radiative correction} 
 
We consider an electron in the field of a nucleus with charge $Z$, i.e. 
\begin{equation*} 
V = - \frac {Z\beta}{|x|}, 
\end{equation*} 
where $\be$ is the \lq\lq real\rq\rq \ fine structure constant $\be = 1/137$. 
The ground state energy of the corresponding Schr\"odinger operator $p^2+ V$ is known to be 
\begin{equation*} 
\infspec [p^2+ V] = -e_0 = - \frac 14 (\be Z)^2. 
\end{equation*} 
The corresponding radiative correction,  obtained in (\ref{rc}), is given by 
\begin{equation}\label{radcorr} 
\Em(V,\la) = - \al 4 (\phi,pD A_V^{-1} p\Da \phi), 
\end{equation} 
where $\phi=\phi(|x|)$ denotes the ground state of $p^2 + V$. 
We know 
\begin{equation}\label{phiz} 
\nabla \phi(|x|) = \partial_r \phi(r) \vec e_r(\theta, \varphi)= e_0^{1/2} \phi(r)  \vec e_r(\theta, \varphi) 
\end{equation} 
when using polar coordinates. 
Denote with $\phi_i$ the eigenstate of $p^2+ V$ with corresponding  eigenvalue $-e_i$. 
Then, by means of (\ref{phiz}), we obtain, by straightforward computations, 
\begin{equation} 
(\phi,pD A_V^{-1} p\Da \phi) \geq 4\pi e_0\sum_{i\geq 1} |c_i|^2 \int_0^\la\frac{p}{e_0 - e_i + p^2 + p} dp \equiv e_0 F(\la), 
\end{equation} 
where 
\begin{equation} 
c_i = \int \overline{\phi_i(r,\theta,\varphi)} \phi(r) \cos(\theta) r^2 dr d \Omega, 
\end{equation} 
with $d \Omega = sin \theta d\theta d\varphi$, and the sum runs over all 
hydrogen eigenstates. 
Notice, due to textbooks $\sum |c_i|^2 \sim 2/15=\frac 23 \frac 15$, 
which in physicists' words is expressed by \lq\lq 80 percent of the average 
excitation energy $\langle e_0 - e_n\rangle_{\rm AV}$ is achieved by the continuous spectrum\rq\rq. \ 
Moreover, 
\begin{equation} 
\ln[1+\la] / \int_0^\la\frac{p}{e_0 - e_i + p^2 + p} dp \to 1 
\end{equation} 
as $\la \to \infty$. That is why, for simplicity, we take 
$\frac{8\pi}{15}\ln[1+\la]$ to evaluate $F(\la)$ 
and obtain an approximative {\it radiative correction} 
\begin{equation}\label{hain} 
R_C = \al e_0\frac{32\pi}{15}  \ln[1+\la] 
\end{equation} 
for the binding energy. (Indeed, for $\la \sim 1$ these two functions perfectly coincide) 
 
\subsection{Calculating concrete values of $\al$ and $\la$} 
 
We search for values of $\al$ and $\la$, that guarantee  the error of the self energy $\err$ being smaller than the 
radiative correction $R_C$. This leads to the condition: 
\begin{equation}\label{alcon} 
\al \leq {\rm Min} \left\{ \frac{\frac {16\pi}{15} e_0 \ln[1+\la]}{\frac{28 
\la^2[1 +\la^2]}{3\pi^2(1-a)} + \frac 7{\pi^2} \la^2}, \frac {a\pi}{4\la}\right\}. 
\end{equation} 
Set $\la = 1/4$ which corresponds to a photon energy $mc^2/2$, that is an 
enormously high value compared to the binding energy $e_0$. 
 
Therefore, the condition on the coupling parameter, such that the radiative correction 
$R_C$ 
dominates the error $\err$, is given by 
\begin{equation} 
\al \leq 0.85 (\beta Z)^2. 
\end{equation} 
Asking for the nuclear charge numbers $Z$ that guarantees 
enhanced binding in the physical case $\al = 1/137$, leads to 
$Z \geq 13$.

However, in the case of smaller values of $\la \sim e_0$, which seems physically 
reasonable, we are in a perfect shape and the error of the self-energy is by far 
negligible compared to the energy shift.

\begin{appendix} 
\section{Auxiliary Lemmas} 
 
\begin{lem}\label{DD} Let $\Psi \in \Hh$. Then 
\begin{multline} 
\Big( \Psi, DD[p^2 + H_f]^{-1} D^*D^*\Psi\Big) \leq \langle 0| 
DD[p^2 + H_f]^{-1} D^*D^*\ora \|\Psi\|^2 \\+ \const \la\Big( 
\big(\Psi,H_f\Psi\big) +  \|p\Psi\|^2\Big). 
\end{multline} 
\end{lem} 
\begin{proof} 
We fix an arbitrary photon number $n$. Recall, 
\begin{multline} 
\F[D^*D^*\psi_n]_{n+2} = \frac 1{\sqrt{(n+2)(n+1)}} 
\sum_{j=1}^{n+2} \sum_{i=1}^{n+1} G(k_j)\cdot G(k_i) 
\times\\\times \psi_n(l,k_1,\dots,\not\!\!  k_j,\dots,\not \!\! 
k_i,\dots,k_{n+2}), 
\end{multline} 
where $\not \!\! k_j$ indicates that the $j-$th variable is omitted. By 
permutational symmetry we can distinguish between three different 
terms, 
\begin{equation} 
\Big( \psi_n, D D [p^2 + H_f]^{-1} D^*D^* \psi_n \Big)= 
I_n+II_n+III_n. 
\end{equation} 
First, the diagonal part, 
\begin{equation} 
I_n= \sum_{\lambda,\nu =1,2} \int \frac{\big[G^\lambda(k_1)\cdot G^\nu(k_2)\big]^2 
|\psi_n(l,k_3,\dots,k_{n+2})|^2}{\big|l-\sum_{i=1}^{n+2}k_i\big|^2 
+ \sum_{i=1}^{n+2}|k_i|} dl dk_1\dots dk_{n+2}. 
\end{equation} 
If we set $\Q =\big|l-\sum_{i=3}^{n+2}k_i\big|^2 + \big|k_1 + 
k_2\big|^2+ \sum_{i=1}^{n+2}|k_i|$ and $b= 
2\big[l-\sum_{i=3}^{n+2}k_i\big]\cdot \big[k_1+k_2\big]$ and use 
the  expansion 
\begin{equation}\label{exp} 
\frac 1{\Q- b} = \frac 1{\Q} + \frac 1{\Q}b\frac 1{\Q} +  \frac 
1{\Q}b\frac 1{\Q-b}b\frac 1{\Q} 
\end{equation} 
then we again see that the second term vanishes when integrating 
over $k_1,k_2$. Therefore, with $\Q \geq \big|k_1 + k_2\big|^2+ 
|k_1| + |k_2|$ we arrive at 
\begin{multline} 
I_n\leq \sum_{\lambda,\nu =1,2}\Big[  \|\Psi\|^2 \int \frac{\big[G^\lambda(k_1)\cdot 
G^\nu(k_2)\big]^2}{|k_1 + k_2|^2 + |k_1| + |k_2|} dk_1 dk_2\\ + 4\int 
\frac{\big[G^\lambda(k_1)\cdot G^\nu(k_2)\big]^2 \big[|k_1| + 
|k_2|\big]^2}{\big[|k_1 + k_2|^2 + |k_1| + 
|k_2|\big]^2(|k_1|+|k_2|)}\times \\ \times 
\big|l-\sum_{i=3}^{n+2}k_i\big|^2|\psi_n(l,k_3,\dots,k_{n+2})|^2dldk_1\dots 
dk_{n+2}\Big] \\ \leq  \langle 0| DD[p^2 + H_f]^{-1} D^*D^*\ora 
\|\psi_n\|^2 +\const \la \|p\psi_n\|^2. 
\end{multline} 
 
For convenience we define the operator $|D|(x)$, which can be 
regarded as the norm of $D(x)$, 
\begin{equation} 
|D|(x) = \sum_{\lambda = 1,2} \int |G^\lambda(k)| e^{ik\cdot x} 
a_\lambda(k) dk 
\end{equation} 
$|D|^*$ denotes the operator adjoint. Similarly, we can define 
$|E|(x)$. Obviously, \cite[Lemma A. 4]{GLL} still holds for the 
\lq\lq norm\rq\rq \ of $D$ and $E$, namely 
\begin{equation}\label{auxeq} 
|D|^*|D| \leq \frac 2\pi H_f, \quad |E|^*|E|\leq \frac 2{3\pi} 
H_f, 
\end{equation} 
that can be proved analogously to (\ref{ddhf}). 
For the second term, with $p^2 + H_f \geq H_f$, we evaluate 
\begin{multline} 
II_n \leq n\sum_{\lambda,\nu = 1,2} \int\frac{\big|G^\lambda(k_1)\cdot G^\nu(k_2)\big|\big|G^\lambda(k_1)\cdot 
G^\nu(k_{n+2})\big|}{\sum_{i=1}^{n+2} |k_i|} \times \\ \times 
|\psi_n(l,k_3,\dots,k_{n+2})||\psi_n(l,k_2,\dots,k_{n+1})| 
dldk_1\dots dk_{n+2}\\ \leq \int \frac{|G(k_1)|^2}{|k_1|}dk_1 
\big(|\psi_n|,|D|^*|D||\psi_n|\big)\leq \const 
\la^2\big(\psi_n,H_f\psi_n\big). 
\end{multline} 
 
Finally, the term where on one side the indices of the created photons differ 
completely from the indices on the other side, 
\begin{multline} 
III_n \leq n^2 \sum_{\lambda,\nu = 1,2}\int\frac{\big|G^\lambda(k_1)\cdot 
G^\nu(k_2)\big|\big|G^\lambda(k_{n+1})\cdot G^\nu(k_{n+2})\big|}{\sum_{i=1}^{n+2} 
|k_i|} \times \\ \times 
|\psi_n(l,k_3,\dots,k_{n+2})||\psi(l,k_1,\dots,k_{n})| dldk_1\dots 
dk_{n+2}\\ \leq \big(|\psi_n|,|D|^*H_f^{-1/2} 
|D|^*|D|H_f^{-1/2}|D||\psi_n|\big) \leq \const \la 
\big(\psi_n,H_f\psi_n\big), 
\end{multline} 
where we used 
\begin{equation} 
\sum_{i=1}^{n+2} |k_i| \geq \left|\sum_{i=1}^{n+1} |k_i| 
\right|^{1/2}\left| \sum_{i=2}^{n+2} |k_i|\right|^{1/2}, 
\end{equation} 
the fact that we can write 
\begin{equation} 
\big[H_f\big]^{-1/2} \psi_n(l,k_1,\dots,k_n) = 
\left[\sum_{i=1}^n |k_i|\right]^{-1/2} \psi_n(l,k_1,\dots,k_n), 
\end{equation} 
and the first equation of (\ref{auxeq}). 
\end{proof} 
 
\begin{lem}\label{off} 
Let $\Psi \in \Hh$ and fix an arbitrary photon number $n$. Then 
\begin{multline} 
\al^{3/2} \big(\sigma E^* \psi_{n+1}, [p^2+H_f]^{-1} \Da\Da \psi_n\big) \leq \\ \leq 
\const \Big[ \al^2 (\la + \la^3) \|\psi_n\|^2 + \al(\la+\la^2) (\psi_{n+1},H_f\pe)\Big]. 
\end{multline} 
\end{lem} 
\begin{proof} 
Obviously, 
\begin{equation}\label{hagi} 
\big(\sigma E^* \psi_{n+1}, [p^2+H_f]^{-1} \Da\Da \psi_n\big)= 
I_n+II_n, 
\end{equation} 
where 
\begin{multline} 
I_n \leq (n+1) \sum_{\lambda,\nu=1,2} \int \frac {\big| H^\lambda(k_1)\big| \big| G^\lambda(k_{1}) \big| 
\big| G^\nu(k_{2})\big|}{|k_1|} \times \\ \times 
|\psi_n(l,k_3,\dots,k_{n+2})||\psi_{n+1}(l,k_2,\dots,k_{n+2})|dldk_1\dots dk_{n+2} 
\\ 
\leq \const \la^2 \|\psi_n\|^{1/2} \int |G(k_2)|\sqrt{|k_2| \rho_{\pe}(k_2)} dk_2 \\ \leq 
\const \big( \al^{1/2}\la^3 \| \psi_n\|^2 +  \al^{-1/2}\la^2 ( \psi_{n+1},H_f \psi_{n+1})  \big) 
\end{multline} 
 
For the second term of (\ref{hagi}) we obtain 
\begin{multline} 
II_n \leq (n+1)^{3/2}\sum_{\lambda,\nu=1,2} \int \frac {\big| H^\lambda(k_1)\big| \big| G^\lambda(k_{n+1}) \big| 
\big| G^\nu(k_{n+2})\big|}{\sum_{i=1}^{n+2} |k_i|} \times \\ \times 
|\psi_n(l,k_1,\dots,k_n)||\psi_{n+1}(l,k_2,\dots,k_{n+2})|dldk_1\dots dk_{n+2} 
\\ 
\leq (n+1)^{3/2}\sum_{\lambda,\nu=1,2}  \int \frac {\big| 
H^\lambda(k_1)\big||\psi_n(l,k_1,\dots,k_n)|}{\left|\sum_{i=1}^{n} |k_i|\right|^{1/2}} 
\times\\ \times \frac{ \big| G^\lambda(k_{n+1}) \big| 
\big| G^\nu(k_{n+2})\big||\psi_{n+1}(l,k_2,\dots,k_{n+2})|}{\left|\sum_{i=2}^{n+1} 
|k_i|\right|^{1/2}}dldk_1\dots dk_{n+2} \\ \leq \||E| H_f^{-1/2} |\psi_n|\|\||D|H_f^{-1/2}|D||\psi_{n+1}|\|, 
\end{multline} 
which implies the statement of the lemma by use of (\ref{auxeq}). 
 
\end{proof}

\end{appendix}

\bigskip 
\noindent {\it Acknowledgment:} The author has been supported by a Marie 
Curie Fellowship 
of the European Community programme \lq\lq Improving Human Research 
Potential and the 
Socio-economic Knowledge Base\rq\rq\ under contract number 
HPMFCT-2000-00660. Furthermore, he thanks his friend Robert Seiringer 
for many valuable discussions and Semjon A. Vugalter for initiating the 
study of this problem.

\end{document}